\documentclass[10pt]{iopart}
\expandafter\let\csname equation*\endcsname\relax
\expandafter\let\csname endequation*\endcsname\relax
\usepackage{amsmath}

\usepackage{psfrag}
\usepackage{graphicx}
\usepackage{graphics}
\usepackage{epsfig}
\usepackage{bm}
\usepackage{color}
\usepackage{verbatim,color}
\usepackage{physics}
\usepackage[acronym]{glossaries}
\usepackage[normalem]{ulem}
\usepackage{mathrsfs}
\usepackage{mathtools}

\newacronym{dr}{DR}{dimensional regularization}
\newacronym{ms}{MS}{minimal subtraction}
\newacronym{eft}{EFT}{effective field theory}
\newacronym{mqst}{MQST}{macroscopic quantum self trapping }
\newacronym{lhy}{LHY}{Lee-Huang-Yang}
\newcommand{\beq}{\begin{equation}}
\newcommand{\eeq}{\end{equation}}
\newcommand{\beqa}{\begin{eqnarray}}
\newcommand{\eeqa}{\end{eqnarray}}
\newcommand{\ba}{\begin{aligned}[b]}
\newcommand{\ea}{\end{aligned}}
\newcommand{\cblue}{\color{black}} 

\newcommand{\draft}{\color{black}}

\begin{document}

\title{Only-phase Popov action: thermodynamic derivation and superconducting electrodynamics}
\author{L. Salasnich$^{1,3,4}$, M.G. Pelizzo$^{2}$, and 
F. Lorenzi$^{1}\footnote{Corresponding author, e-mail: francesco.lorenzi.2@phd.unipd.it}$}
\address{$^{1}$Dipartimento di Fisica e Astronomia 
``Galileo Galilei", Universita di Padova, and INFN Sezione di Padova, 
Via Marzolo 8, 35131, Padova, Italy
\\
$^{2}$ 
Dipartimento di Ingegneria dell’Informazione, Università di Padova, via Gradenigo, 6B, 35131 Padua, Italy, and Istituto di Fotonica e Nanotecnologie, Consiglio Nazionale delle Ricerche, via Trasea 7, 35131 Padua, Italy
\\
$^{3}$Padua Quantum Technologies Research Center, Universita di Padova, 
Via Gradenigo 6B 35131, Padova, Italy
\\
$^{4}$Istituto Nazionale di Ottica del Consiglio Nazionale delle Ricerche,  
Via Nello Carrara 2, 50019 Sesto Fiorentino, Italy }

\begin{abstract}
We provide a thermodynamic derivation of the only-phase Popov 
action functional, which is often adopted to study the low-energy 
effective hydrodynamics of a generic {nonrelativistic} superfluid. 
It is shown that the crucial assumption is the use of the 
saddle point approximation after neglecting the quantum-pressure term. 
As an application, we analyze charged superfluids (superconductors) 
coupled to the electromagnetic field at zero temperature. 
Our only-phase and minimally-coupled theory predicts 
the decay of the electrostatic field inside a superconductor with a characteristic length much smaller than the London penetration depth of the static magnetic field. {\cblue This result is confirmed also by a relativistic only-phase Popov action we obtain from the Klein-Gordon Lagrangian.}  

\end{abstract}

\maketitle

\section{Introduction}
The phenomenological description of superconductive hydrodynamics has gained recent interest, due to the possibility of writing effective hydrodynamic Lagrangians describing the interaction of electromagnetic fields with charged superfluids \cite{hirsch2004, hirsch2015, grigorishin2021}. In this context, the only-phase Popov action is a valuable tool {  to predict} the behavior of the Nambu-Goldstone phase field \cite{nambu1960, goldstone1961}. 
On the other hand, the problem of the screening of the electromagnetic field inside a superconductor has been deeply investigated since the first phenomenological models \cite{london1935, london1936}, where it was shown that the magnetostatic field is exponentially screened with a characteristic length called London penetration depth $\lambda_L$. Typically $\lambda_L$ is in the order of hundreds of nanometers. Past theories suggest that the same cannot be said for the electrostatic field screening, that, analogously to the case of normal conductors, is decaying with a much shorter length scale, i.e. the Thomas-Fermi screening length, in the order of few angstr\"oms \cite{mermin-book}.

In the present work, on a theoretical point of view, we first review the Popov prescription for obtaining an only-phase action for a nonrelativistic zero-temperature fluid \cite{popov-book} in Section \ref{sec:2popov}. Then, in Section \ref{sec:3popov} we show how it is possible to derive the same result from the familiar hydrodynamic action of a self-interacting nonrelativistic bosonic field. We observe that the result by Popov can be related to the latter action, given that the quantum pressure term has been neglected. By introducing a path integration over the number density field and an additional field, and performing a saddle-point approximation of the grand-canonical partition function, the original prescription by Popov for obtaining the only-phase action is retrieved.
Working at zero temperature, we identify the additional field as the Nambu-Goldstone field, that appears in the resulting Lagrangian density in its gradient squared.
In Section \ref{sec:4super} we introduce minimal coupling of the Nambu-Goldstone phase field to the electromagnetic field, and we obtain the equations of motion including one-loop corrections of the superconductive dynamics. They involve the Maxwell equations, and constitutive relations for the charge density and current density. Comparing the penetration depth of the magnetostatic field to the electrostatic field, we conclude that, within our formalism, the electrostatic field penetration depth can be put on the same footing as the magnetostatic field one, {  but with a penetration depth $\lambda_E$ many orders of magnitude smaller than  $\lambda_L$}. 
{\cblue Finally, in Section \ref{sec:5super} we develop a relativistic only-phase Popov model  which confirms that $\lambda_E\ll \lambda_L$ 
in realistic superconductors.} 
Experiments measuring the penetration depth in superconductors are hindered by the necessity to use, at the same time, precise field measurements with nanometer-scale resolution, and cryogenic apparatus to keep the superconductor well below the transition temperature \cite{peronio2016}. We remark that, until now, experiments have not been able to measure the electrostatic field penetration depth to a sufficient accuracy to discriminate between the prediction of the present work, i.e. the penetration depth indicated by {  $\lambda_E$}, and the Thomas-Fermi screening effect.

\section{Popov superfluid Lagrangian} \label{sec:2popov}

In the grand canonical framework, at zero temperature 
the pressure $P$ of a fluid can be written in terms of its chemical 
potential $\mu$, i.e. 
\beq 
P=P(\mu) \; . 
\eeq
This is the zero-temperature equation of the state of 
the fluid \cite{bec-book,landau}.  { For instance, in the case of a weakly-interacting bosonic gas it is given by $P(\mu)=\mu^2/(2g)$, where $g$ is the strength of the effective Bose-Bose contact interaction. Instead, for a two-spin-component superfluid Fermi gas one has $P(\mu)=(2/(15\pi^2))(2m/\hbar^2)^{3/2}\mu^{5/2}$ neglecting the Fermi-Fermi interaction.} The number density $n$ can be obtained from the pressure $P(\mu)$ 
using the thermodynamic formula \cite{landau}
\beq 
n = {\partial P \over \partial \mu}(\mu) \: . 
\label{density}
\eeq
For a fluid of identical particles of mass $m$ and chemical 
potential $\mu(n)$, the zero-temperature speed of sound {  $c_s$} 
is defined as \cite{landau}
\beq 
{  c_s} = \sqrt{{n\over m} {1\over {\partial^2P\over \partial \mu^2}(\mu)}} \; . 
\label{sound}
\eeq

The main idea of Popov \cite{popov1972,popov-book}, later adopted and extended 
by other authors (see, for instance, \cite{witten1989,son2006,son2012}), 
is that the only-phase action functional 
\beq 
\tilde{S}[\theta] = \int \dd{t} \int \dd[D]{\bf r} \, \tilde{\mathscr{L}} 
\label{final}
\eeq
of a nonrelativistic superfluid, which is 
characterized by the Naubu-Goldstone \cite{nambu1960,goldstone1961} 
real scalar field $\theta({\bf r},t)$, is obtained with the prescription 
\beq 
\mu \to \mu - \hbar {\partial_t \theta} - {\hbar^2 \over 2m} 
|{{\boldsymbol\nabla}\theta}|^2
\label{prescrivo}
\eeq
into the pressure $P(\mu)$ such that 
\beq 
\tilde{\mathscr{L}} 
= P\left(\mu - \hbar {\partial_t \theta} - {\hbar^2 \over 2m} 
|{{\boldsymbol\nabla}\theta}|^2\right) \;  
\label{magic0}
\eeq 
is the real-time only-phase Lagrangian density. This approach has been 
also extended to the relativistic case 
\cite{rattazzi2012,nicolis2013,nicolis2019,torrieri2019,vari2022}.

Expanding (\ref{magic0}) with respect to $\theta$ around $\mu$, 
taking into account Eqs. (\ref{density}) and (\ref{sound}), we find 
\beq 
\tilde{\mathscr{L}} = P(\mu) - n \, 
( \hbar {\partial_t \theta} + {\hbar^2 \over 2m} 
|{{\boldsymbol\nabla}\theta}|^2) 
+ {1\over 2} {n\over m{  c_s}^2} \, 
(\hbar{\partial_t \theta} + {\hbar^2 \over 2m} 
|{{\boldsymbol\nabla}\theta}|^2)^2 + ... 
\label{dottica}
\eeq
Removing the dots ($...$), Eq. (\ref{dottica}) becomes 
exactly the zero-temperature 
low-wavenumber effective Lagrangian density one finds at the one-loop level 
from the microscopic {  beyond-mean-field BCS-like} model of attractive fermions \cite{schakel1990} 
and also from the microscopic model of weakly-interacting 
bosons \cite{schakel1994}. Instead, considering only the first two 
terms of Eq. (\ref{dottica}) one recovers 
the familiar hydrodynamic Lagrangian density 
\beq 
\tilde{\mathscr{L}}_0 
= P(\mu) - n \, ( \hbar {\partial_t \theta} + {\hbar^2 \over 2m} 
|{{\boldsymbol\nabla}\theta}|^2)
\label{familiar}
\eeq
of classical inviscid and irrotational fluids 
(see, for instance, \cite{sala2013}). 

\section{Deriving the only-phase Popov action} \label{sec:3popov}

The derivation of the only-phase Popov action, Eq. (\ref{magic0}), 
was performed by Popov \cite{popov1972,popov-book} starting from a bosonic 
action and separating fast and slowly varying components of the bosonic 
field. Here we obtain the same result by using a different 
procedure: the saddle-point functional integration over the density field 
of a peculiar density-phase action functional, given by 
\beq 
S[n,\theta] = \int_0^{+\infty} 
\dd{t} \int_{L^D} \dd[D]{\bf r} \, [ - {\cal E}_0(n) - n \, \hbar \partial_t \theta 
- n \, {\hbar^2\over 2m} |{\boldsymbol \nabla} \theta|^2 + \mu \, n ] \; ,  
\label{initial0}
\eeq
where ${\cal E}_0(n)$ is the zero-temperature internal energy of the system 
as a fuction of the local number density $n({\bf r},t)$. 
{ For instance, in the case of weakly-interacting bosons 
${\cal E}_0(n)=gn^2/2$ 
with $g$ the interaction strength, while for superfluid fermions ${\cal E}_0(n)=(5/3)(3\pi^2)^{2/3}n^{5/3}$ again neglecting the residual inter-particle interaction between fermions.} 
Eq. (\ref{initial0}) is nothing else than the hydrodynamic action of a self-interacting 
nonrelativistic bosonic field  $\psi({\bf r},t)$ characterized by the 
action functional
\beq 
S[\psi] =  \int_0^{+\infty} \dd{t} \int_{L^D} \dd[D]{\bf r} 
\left[ i{\hbar\over 2} (\psi^* \partial_t \psi - \psi \partial_t \psi^*) 
- {\hbar^2\over 2m}|{\boldsymbol\nabla}\psi|^2 - {\cal E}_0(|\psi|^2) 
\right] \; , 
\eeq
under the familiar Madelung decomposition 
\beq 
\psi({\bf r},t) = \sqrt{n({\bf r},t)} \, e^{i(\theta({\bf r},t)-{\mu\over\hbar} t)}
\label{mad-lung}
\eeq
but then neglecting the quantum pressure term 
$\hbar^2(\nabla \sqrt{n})^2/(2m)$. 
{
  This assumption is reliable in the spatial regions where the condition $ | \hbar^2 (\nabla \sqrt{n})^2 /(2m) | \ll |{\cal E}_0(n)|$ is satisfied. The inequality says that the quantum pressure term can be neglected if the local number density is much larger than its gradient. Usually, the quantum pressure term is relevant, in the presence of a confinement potential, only near the surface or, in the present case, only at very small wavelengths.}

\subsection{Grand Canonical  partition function, free energy and 
grand potential}

It is well known that the Grand Canonical partition function ${\cal Z}$, 
that is a function of the chemical potential $\mu$, is related 
to the Helmholtz free energy $F$, that is a function of the total 
number $N$ of particle, by the thermodynamic formula 
\beq 
{\cal Z} = \sum_{N=0}^{\infty} e^{-\beta [ F(N) -\mu N ]} \; ,  
\eeq
where $\beta=1/(k_BT)$ with $k_B$ the Boltzmann constant and $T$ the 
absolute temperature. 
{  Remember that the Helmholtz free energy $F$ is d{\cblue e}fined as $F={E}-T S$ with $E$ the internal energy and $S$ the entropy.}

{  In the low-temperature regime, where $\beta$ becomes very large, one can adopt the saddle-point approximation finding}  
\beq 
{\cal Z} \simeq e^{-\beta [ F(N_{s}) -\mu N_{s} ]} \; , 
\label{saddle1}
\eeq
where $N_{s}$ is the saddle-point number of particles, 
obtained by inverting the formula 
{  
\beq 
{\partial \over \partial N} [F(N)-\mu N] = 0 \; ,  
\eeq
}
which extremizes the exponent of the exponential function. 
It is important to stress that $N_{s}$ is a function of the chemical 
potential $\mu$, i.e. $N_{s}=N_{s}(\mu)$. Thus, we can write 
\beq 
{\cal Z} \simeq e^{-\beta \Omega(\mu)} \; , 
\label{saddle2}
\eeq
where $\Omega(\mu)$ is the thermodynamic grand potential, such that 
\beq 
\Omega(\mu) = F(N_{s}(\mu)) - \mu N_{s}(\mu) \; . 
\eeq

\subsection{Path-integral representation adding an arbitrary field}

Making explicit the procedure briefly discussed in Ref. \cite{babaev-book}, 
let us introduce the number density field $n({\bf r},\tau)$ as a 
function of the position vector ${\bf r}$ and imaginary 
time $\tau$. It must satisfy the relation 
\beq 
\beta N = {1\over \hbar} \int_0^{\hbar\beta} \dd{\tau} \int_{L^D} \dd[D]{\bf r} \, 
n({\bf r},\tau) \; . 
\eeq
We also introduce the local free energy density 
${\cal F}(n({\bf r},t),\theta({\bf r},t))$ which depends on the 
local number density $n({\bf r},\tau)$ and another generic field 
$\theta({\bf r},t)$. We impose that 
\beq 
\beta F = {1\over \hbar} \int_0^{\hbar\beta} \dd{\tau} \int_{L^D} \dd[D]{\bf r} \, 
{\cal F}(n({\bf r},\tau),\theta({\bf r},\tau)) \; . 
\eeq 
Then, taking into account {  the thermodynamic limit and the fact that the particle number density is not uniform}, we write the relationship {  (see also \cite{babaev-book})}
\beq 
\sum_{N=0}^{\infty} {  \rightarrow \int_0^\infty \dd N} \rightarrow \int {\cal D}[n({\bf r},t)] \; ,
\eeq
{  immediately obtaining} the following path-integral representation 
of the Grand Canonical partition function 
\beq 
{\cal Z}[\theta] =  \int {\cal D}[n({\bf r},\tau)] \,  
e^{-{1\over\hbar} \int_0^{\hbar\beta} 
\dd{\tau} \int_{L^D} \dd[D]{\bf r} \, \left[ 
{\cal F}(n({\bf r},\tau),\theta({\bf r},\tau)) 
-\mu \, n({\bf r},\tau) \right]} \; . 
\label{mysterious}
\eeq

We use also here the saddle-point approximation. In this way, we have 
\beq 
{\cal Z}[\theta] \simeq 
e^{-{1\over\hbar} \int_0^{\hbar\beta} 
\dd{\tau} \int_{L^D} \dd[D]{\bf r} \, \left[ 
{\cal F}(n_s({\bf r},\tau),\theta({\bf r},\tau)) 
-\mu \, n_s({\bf r},\tau) \right]} \; , 
\eeq
where the saddle-point density field $n_s({\bf r},\tau)$ is obtained 
by inverting the equation 
\beq
\mu = {\delta {\cal F}\over \delta n}(n_s,\theta) 
\eeq
which involves the functional derivative of the local free energy. 
Clearly, this saddle-point density $n_{s}({\bf r},\tau)$ is a function of the 
chemical potential $\mu$, i.e. $n_{s}({\bf r},\tau;\mu)$. 

We introduce a local pressure ${P}(\theta({\bf r},\tau);\mu)$ 
that is a function of the arbitrary field $\theta({\bf r},t)$ and also 
of the chemical potential $\mu$. This local pressure, that is given by 
\beq 
{P}(\theta({\bf r},\tau);\mu) =  - 
{\cal F}(n_s({\bf r},\tau;\mu),\theta({\bf r},\tau)) 
+\mu \, n_s({\bf r},\tau;\mu) \; , 
\eeq
is related to the grand potential by the formula 
\beq 
-\beta \Omega = {1\over \hbar} 
\int_0^{\hbar\beta} \dd{\tau} \int_{L^D} \dd[D]{\bf r} \, 
{P}(\theta({\bf r},\tau);\mu) \; . 
\eeq

\subsection{Zero temperature limit}

In the zero-temperature limit, i.e. setting $\beta\to +\infty$, 
where the local free energy density ${\cal F}(n,\theta)$ becomes a local 
internal energy density ${\cal E}(n,\theta)$ {  because the entropic contribution $T S$ vanishes}, 
and performing the Wick rotation 
\beq 
\tau = i \, t \; , 
\eeq
from Eq. (\ref{mysterious}) we get 
\beq 
e^{{i\over\hbar} {\tilde S}[\theta]} = 
e^{{i\over\hbar} S[n_s(\mu),\theta]} \simeq  \int {\cal D}[n] \,  
e^{{i\over\hbar} S[n,\theta]} \; , 
\label{mainornot}
\eeq
where  
\beq 
{\tilde S}[\theta] = \int_0^{+\infty} 
\dd{t} \int_{L^D} \dd[D]{\bf r} \, {P}(\theta({\bf r},t);\mu)
\label{myst1}
\eeq
is the action functional without the local density. Instead, 
\beq 
S[n,\theta] = \int_0^{+\infty} \dd{t} \int_{L^D} \dd[D]{\bf r} \, 
\left[ - {\cal E}(n({\bf r},t),\theta({\bf r},t)) 
+ \mu \, n({\bf r},t) \right] 
\label{initial}
\eeq
is the action functional with the local density. {  The last equality of Eq. (\ref{mainornot})  
is the zero-temperature version of Eqs. (\ref{saddle1}) and (\ref{saddle2}).} 

\subsection{Nambu-Goldstone phase field} 

Let us suppose that the field $\theta({\bf r},t)$ is the Nambu-Goldstone 
phase field of a superfluid \cite{nambu1960,goldstone1961}, such that 
\beq 
{\bf v}_s = {\hbar\over m} {\boldsymbol\nabla} \theta \; 
\label{superv}
\eeq
is the superfluid velocity of the system composed of identical 
bosonic (or bosonic-like) particles of mass $m$. Here $\hbar$ is the 
reduced Planck constant. Eq. (\ref{initial}) is the 
density-phase Popov action imposing that 
the internal energy density 
${\cal E}(n({\bf r},t),\theta({\bf r},t))$ is given by 
\beq 
{\cal E}(n,\theta) = {\cal E}_0(n) 
+ n \hbar \partial_t \theta + n {\hbar^2\over 2m} |{\boldsymbol \nabla} \theta|^2 \; , 
\label{assum}
\eeq
where ${\cal E}_0(n)={\cal E}(n,\theta=0)$ is 
the zero-temperature internal energy 
in the absence of the phase field $\theta({\bf r},t)$. 

At zero temperature, the thermodynamic 
formula which connects the saddle-point local density $n_s({\bf r},t)$ 
to the chemical potential $\mu$ is 
\beq 
\mu = {\partial {\cal E}\over \partial n}(n_s,\theta) \; . 
\eeq
Explicitly, we have 
\beq 
\mu =  {\partial {\cal E}_0\over \partial n}(n_s) + 
\hbar \partial_t \theta + {\hbar^2\over 2m} |{\boldsymbol \nabla} \theta|^2 \; 
\eeq
or, equivalently 
\beq
\mu - \hbar \partial_t \theta - {\hbar^2\over 2m} |{\boldsymbol \nabla} \theta|^2
=  {\partial {\cal E}_0\over \partial n}(n_s) \; . 
\label{pao}
\eeq 

The inversion of Eq. (\ref{pao}), namely 
\beq 
n_s = n_s(\mu,\theta) = \left({\partial {\cal E}_0\over \partial n}\right)^{-1}
(\mu - \hbar \partial_t \theta - {\hbar^2\over 2m} |{\boldsymbol \nabla} \theta|^2) \; , 
\eeq
gives $n_s$ as a function of $\mu$ and $\theta$. In this way we can 
then write the formal expression 
\beqa 
{P}(\mu,\theta) &=& P(n_s(\mu,\theta)) = 
P\left(\left({\partial {\cal E}_0\over \partial n}\right)^{-1}
( 
\mu - \hbar \partial_t \theta - {\hbar^2\over 2m} |{\boldsymbol \nabla} \theta|^2)\right)
\nonumber 
\\
&=& {P}\left(\mu - \hbar \partial_t \theta - {\hbar^2\over 2m} |{\boldsymbol \nabla} \theta|^2\right) \; ,
\label{onthebasis}
\eeqa
that is a Legendre transformation and ${P}(\mu,\theta)$ is the 
local pressure which appears in Eq. (\ref{myst1}). 

It is important to observe that in Eq. (\ref{myst1}) there is the 
peculiar Lagrangian density 
\beq 
\tilde{\mathscr{L}} = { P}(\mu,\theta) \; .  
\label{mysto}
\eeq
Clearly, if $\theta({\bf r},t)=0$, the Lagrangian density is nothing else 
than the zero-temperature pressure $P$ written in terms of its chemical 
potential $\mu$, i.e. $P(\mu)={P}(\mu,\theta=0)$. 
On the basis of Eq. (\ref{onthebasis}), 
the Lagrangian (\ref{mysto}) is given exactly by Eq. (\ref{magic0}), 
thus ${P}(\mu,\theta)={ P}(\mu-\hbar {\partial_t \theta} - 
\hbar^2|{{\boldsymbol\nabla}\theta}|^2/(2m))$. 

\section{Superconducting Lagrangian} \label{sec:4super}

In the case of a superconductor, i.e. a charged superfluid 
with $q$ the electric charge of each particle of mass $m$, 
one can generalize the Lagrangian density (\ref{magic0}) 
introducing the following coupling \cite{son2006,babaev-book,schakel-book}
\beqa 
{\partial_t}\theta &\to& {\partial_t}\theta + {q\over \hbar} \Phi
\label{schabo1}
\\
{\boldsymbol \nabla}\theta &\to& {\boldsymbol \nabla}\theta 
- {q\over \hbar} {\bf A}
\label{schabo2}
\eeqa
to the electromagnetic field. Here $\Phi({\bf r},t)$ is the scalar potential 
and ${\bf A}({\bf r},t)$ is the vector potential, such that 
\beqa
{\bf E} &=& - {\boldsymbol\nabla}\Phi - \partial_t {\bf A} 
\label{eanda}
\\
{\bf B} &=& {\boldsymbol \nabla} \wedge {\bf A} 
\label{banda}
\eeqa
with ${\bf E}({\bf r},t)$ the electric field and ${\bf B}({\bf r},t)$ 
the magnetic field. In this way, the total Lagrangian density 
$\mathscr{L}_{\rm tot}$ of the Goldstone mode coupled to the electromagnetic 
field is given by 
\beq 
\mathscr{L}_{\rm tot} = \mathscr{L}_{\rm{shift}} + \mathscr{L}_{\rm{em}}+\mathscr{L}_{\rm{bg}} \; , 
\label{kari}
\eeq
where 
\beq 
\mathscr{L}_{\rm{shift}} = 
{P}\left(\mu - \hbar (\partial_t\theta + {q\over \hbar} 
\Phi ) - {\hbar^2 \over 2m} 
|{\boldsymbol \nabla}\theta - {q\over \hbar} {\bf A}|^2\right)   
\label{kari1}
\eeq
is the Lagrangian density of the shifted Goldstone mode, 
\beq 
\mathscr{L}_{\rm{em}} = {\epsilon_0\over 2} |{\bf E}|^2 
- {1\over 2\mu_0} |{\bf B}|^2  
\label{kari2}
\eeq
is the Lagrangian density of the free electromagnetic field, with 
$\epsilon_0$ the dielectric constant in the vacuum and $\mu_0$ the 
paramagnetic constant in the vacuum. Remember that $c=1/\sqrt{\epsilon_0\mu_0}$ is the speed of light in the vacuum. 
We also added a term 
\beq
\mathscr{L}_{\rm{bg}} = \bar{n}_{\rm{bg}}q \Phi
\eeq
that takes into account the role of a uniform background of positive charges, i.e. the average number density of the ions $\bar{n}_{\rm{bg}}$ times the electric charge $q$, to ensure net neutrality of the material, similarly to the Jellium model of a conductor. 

The Euler-Lagrange equations of the total Lagrangian (\ref{kari}) 
with respect to the scalar potential $\Phi({\bf r},t)$ and the vector 
potential ${\bf A}({\bf r},t)$ are nothing else than the 
Maxwell equations 
\beqa
{\boldsymbol\nabla} \cdot {\bf E} &=& {\rho\over \epsilon_0} 
\label{max1}
\\
{\boldsymbol\nabla} \cdot {\bf B} &=& 0 
\label{max2}
\\
{\boldsymbol\nabla} \wedge {\bf E} &=& {\partial_t {\bf B}}
\label{max3}
\\
{\boldsymbol\nabla} \wedge {\bf B} &=& \mu_0 \ {\bf j} + 
{\epsilon_0 \mu_0} \ \partial_t{\bf E} 
\label{max4}
\eeqa
where, however, the expressions of the local charge 
density $\rho({\bf r},t)$, including Cooper pairs and the uniform positive background, and the local current density ${\bf j}({\bf r},t)$ 
are highly nontrivial 
\beqa
{\rho} &=& -{\partial ({\cal L}_{\rm shift} + {\cal L}_{\rm bg}) \over \partial \Phi} 
\\
{\bf j} &=&  - {\partial {\cal L}_{\rm shift}\over \partial {\bf A}}  
\eeqa
Notice that, within the approximation of using the 
Lagrangian (\ref{familiar}) instead of (\ref{magic0}), one gets 
\beqa
{\rho} &=& q n - q \bar{n}_{\rm bg}
\\
{\bf j} &=& q n {\bf v}_s - {q^2 n\over m} {\bf A}  
\eeqa
where the second term in the current density is nothing else than 
the London current \cite{london1935}, which gives rise to 
the expulsion of a magnetic field from a superconductor 
(Meissner-Ochsenfeld effect) \cite{meissner1933}. 

With an improved approximation, namely working with the 
expansion (\ref{dottica}) {  and including next-to-leading terms}, we find instead 
\beqa
{\rho} &=& q \, n - q \bar{n}_{\rm bg} - \epsilon_0 \, {q^2 n \mu_0\over m} 
{  
\left({c^2\over {  c_s}^2}\right)} \, \Phi  
\label{gulp1}
\\
{\bf j} &=&  q \, n {\bf v}_s - {q^2 n \over m} 
\, {\bf A}  \;  \label{gulp2}
\eeqa
{  taking into account Eq. (\ref{sound}) which gives $P''(\mu)=n/(m{  c_s}^2)=\epsilon_0\mu_0 (n/m)(c/c_s)^2$ 
with $c=1/\sqrt{\epsilon_0\mu_0}$ the speed of light and $c_s$ the speed of sound}. 
It is important to stress that $q n$ 
is the electric charge density of Cooper pairs, $- q \bar{n}_{\rm bg}$ is the electric charge density of the uniform background, and 
$-\epsilon_0\mu_0 q^2 n\Phi/m$ 
is {  a sort of} interaction charge density related to the coupling 
with the scalar potential $\Phi$. {  In Section 4.2 we will show that this term is crucial to get the correct penetration depth of the electric field.} Instead, $q \, n {\bf v}_s$ is the electric current density of Cooper pairs 
and $-{q^2 n}{\bf A}/m$ is the London current \cite{london1935} 
related to the coupling with the vector potential ${\bf A}$. {\draft We remark that {  a} result {  similar to the one of} Eq.~(\ref{gulp1}) can be obtained from the relativistic equation of motion for the Cooper pair field, i.e. Klein-Gordon equation, by coupling the equation to the electromagnetic field, and taking the nonrelativistic limit \cite{sala2023}. However, within that mean-field relativistic approach \cite{sala2023} the ratio $(c^2/c_s^2)$ does not appear in the last term of Eq. (\ref{gulp1}). 

\subsection{London penetration depth for the magnetostatic field}

In a static configuration with a zero superfluid velocity ${\bf v}_s$ 
and in the absence of the electric field, i.e. 
${\bf E}={\bf 0}$, the curl of Eq. (\ref{max4}) gives 
\beq 
- \nabla^2 {\bf B} = \mu_0 \ {\boldsymbol \nabla} 
\wedge \left( - {q^2 n\over m} {\bf A} \right)  \; , 
\label{crucial}
\eeq
taking into account that ${\boldsymbol\nabla} \wedge ({\boldsymbol \nabla} \wedge {\bf B})= 
- \nabla^2 {\bf B} + {\boldsymbol \nabla} ({\boldsymbol \nabla} \cdot 
{\bf B}) = - \nabla^2 {\bf B}$ due to the Gauss law, Eq. (\ref{max2}). Assuming that 
the local density $n({\bf r})$ is uniform, i.e. $n({\bf r})={\bar n}$, 
by using Eq. (\ref{banda}) we get 
\beq 
\nabla^2 {\bf B} = {q^2 {\bar n}_s \mu_0 \over m} {\bf B} \; .
\eeq
Choosing the magnetic field as ${\bf B} = B(x)\, {\bf u}$, with ${\bf u}$ 
a unit vector, the previous equation 
has the following physically relevant solutions for 
a superconducting slab defined in the region $x\geq 0$:  
\beq 
B(x) = B(0) \ e^{-x/\lambda_L} \; , 
\label{mah}
\eeq
where 
\beq 
\lambda_L = \sqrt{m\over q^2 \bar{n}_s \mu_0} 
\label{lambdaL}
\eeq
is the so-called London penetration depth \cite{london1935}, which is 
around $100$ nanometers \cite{annett-book}. 
Eq. (\ref{mah}) says that inside a superconductor 
the magnetostatic field decays exponentially. This is a well-known 
Meissner-Ochsenfeld effect \cite{meissner1933}. 

\subsection{Penetration depth for the electrostatic field}

It is well known that normal metals screen an external electric 
field ${\bf E}$, which can penetrate at most few 
angstr\"oms (Thomas-Fermi screening length) \cite{mermin-book}. 
For superconducting materials, our equations (\ref{max1}), (\ref{max2}), 
(\ref{gulp1}), and (\ref{gulp2}) suggest that the electric 
field ${\bf E}$ exponentially decays inside a zero-temperature 
superconductor with {  a characteristic}   penetration depth
{  
\beq
\lambda_E = 
\lambda_L {c_s\over c} 
\label{lambdaE}
\eeq
which is many orders of magnitude much smaller than the London penetration depth $\lambda_L$.} 
Let us show how to derive this result within our theoretical 
framework. In a static configuration, in the absence the  
magnetic field, i.e. {\bf B}={\bf 0}, and assuming a uniform 
number density, the gradient of Eq. (\ref{max1}), with Eq. (\ref{gulp1}) 
and Eq. (\ref{lambdaL}), gives 
\beq 
\nabla^2 {\bf E} = - {1\over {  \lambda_E}^2} {\boldsymbol\nabla} \Phi  \; , 
\eeq
taking into account that ${\boldsymbol\nabla} ({\boldsymbol \nabla} \cdot {\bf E})= 
\nabla^2 {\bf E} - {\boldsymbol \nabla} \wedge 
({\boldsymbol \nabla} \wedge {\bf E}) = \nabla^2 {\bf E} \; $.
In addition, Eq. (\ref{eanda}) {  with the $\partial_t {\bf A}={\bf 0}$ implies ${\bf E}=-{\boldsymbol\nabla} \Phi$ and consequently we find}
\beq 
\nabla^2 {\bf E} = {1\over {  \lambda_E}^2} {\bf E}  \; .  
\eeq
Choosing ${\bf E} = E(x)\, {\bf u}$, with ${\bf u}$ a unit vector, 
the previous equation 
has the following physically relevant solutions for 
a superconducting slab defined in the region $x\geq 0$: 
\beq 
E(x) = E(0) \ e^{-x/{  \lambda_E}} \; . 
\label{rimah}
\eeq
Eq. (\ref{rimah}) says that inside a zero-temperature 
superconductor the electrostatic field decays exponentially. 
The characteristic decay length {  $\lambda_E$} of the electric field 
is {  quite different} with respect to the London penetration depth $\lambda_L$ of the magnetic field. 
Eq. (\ref{rimah}) was predicted by the London brothers \cite{london1935} 
{  with $\lambda_L$ instead of our $\lambda_E$} 
but, in the absence of experimental validation \cite{london1936}, 
subsequently, Fritz London rejected { it} \cite{london-book}. 

{\cblue 

\section{Relativistic only-phase Popov action} 
\label{sec:5super}

On the basis of the procedure previously discussed, it is possible to obtain a rel  oativistic only-phase Popov action. Following Ref. \cite{sala2023} we start from the relativistic Klein-Gordon 
complex scalar field $\varphi({\bf r},t)$ with Lagrangian density 
\beq
\mathscr{L}_{\rm R} = {\hbar^2\over 2mc^2} |\partial_t\varphi|^2 - 
        {\hbar^2\over 2m} |{\boldsymbol\nabla}\varphi|^2
- {mc^2\over 2} |\varphi|^2 
        - {\cal E}_0(|\varphi|^2)  + \mu {i\hbar^2\over 2m c^2} \left( 
\varphi^* \partial_t\varphi 
- \varphi \partial_t \varphi^*
\right) \; . 
\label{lstart}
\eeq
where the last term takes into 
account the conserved quantity 
\beq 
Q = {i\hbar^2\over 2m c^2} 
\int d^3{\bf r} \left( 
\varphi^* \partial_t\varphi 
- \varphi \partial_t \varphi^*
\right)
\eeq
that is the number of particles minus the number of anti-particles \cite{kapusta-book}. 
This Lagrangian can be rewritten in a Schr\"odinger-like form setting 
\beq 
\varphi({\bf r},t) = \psi({\bf r},t) \, e^{-imc^2t/\hbar} \; 
\eeq
with the aim of removing the mass term 
$mc^2|\varphi|^2/2$. 
In this way we get 
\beqa 
\mathscr{L}_{\rm R} &=& {\hbar^2\over 2mc^2}|\partial_t\psi|^2 +
        {i\hbar\over 2} 
(\psi^*\partial_t\psi - \psi\partial_t\psi^*) 
- {\hbar^2\over 2m} |{\boldsymbol\nabla}\psi|^2 - {\cal E}_0(|\psi|^2) 
\nonumber 
\\
&+& \mu \left[ |\psi|^2 
+ {i\hbar\over 2mc^2} 
\left(\psi^*\partial_t\psi 
-  \psi\partial_t\psi^*\right)
\right]
\;  .
\eeqa
The terms with $mc^2$ at the denominator make the relativistic Lagrangian different with respect to the nonrelativistic one. We now insert 
\beq 
\psi({\bf r},t) = \sqrt{n({\bf r},t)} \, 
e^{i \theta({\bf r},t)}
\eeq
into the last Lagrangian density obtaining
\beq
\mathscr{L}_{\rm R} =  n \, {\hbar^2\over 2mc^2} (\partial_t \theta)^2 
 - n \, \hbar \partial_t \theta 
- n \, {\hbar^2\over 2m} |{\boldsymbol \nabla}\theta|^2 - {\cal E}_0(n) 
+ \mu \, n 
\label{popov-relativistic}
\eeq
after neglecting the terms that depend on the
space and time derivatives of the density $n({\bf r},t)$, i.e. $\hbar^2(\nabla \sqrt{n})^2/(2m)$ and $\hbar^2(\partial_t\sqrt{n})^2/(2m)$, and also the direct coupling between $\mu$ and $\partial_t\theta$. Eq. (\ref{popov-relativistic}) is a density-phase Popov Lagrangian 
with a relativistic correction, i.e. the term $n\hbar^2(\partial_t\theta)^2/(2mc^2)$.
Then, as a direct consequence of the  Legendre transformation 
discussed in Section \ref{sec:3popov}: 
\beq 
{\tilde S}_{\rm R}[\theta] = \int_0^{+\infty} dt \int_{L^D} d^D{\bf r} \  \tilde{\mathscr{L}}_{\rm R} 
\eeq
with 
\beq 
\tilde{\mathscr{L}}_{\rm R} 
= {P}\left(
{\mu 
+{\hbar^2\over 2mc^2} (\partial_t \theta)^2  - 
\hbar {\partial_t \theta} - {\hbar^2 \over 2m} 
|{{\boldsymbol\nabla}\theta}|^2}
\right) \;  
\label{magic0-relativistic} \;  
\eeq  
the relativistic only-phase Popov Lagrangian density. 
Please, compare it with the non-relativistic one, Eq. (\ref{magic0}). Expanding (\ref{magic0-relativistic}) with respect to $\theta$ around $\mu$, we find 
\beqa 
\tilde{\mathscr{L}}_{\rm R} &=& P(\mu) - n \, 
( - {\hbar^2\over 2mc^2} (\partial_t \theta)^2 +
\hbar {\partial_t \theta} + {\hbar^2 \over 2m} 
|{{\boldsymbol\nabla}\theta}|^2) 
\nonumber
\\
&+& {1\over 2} {n\over m{  c_s}^2} \, 
(-{\hbar^2\over 2mc^2} (\partial_t \theta)^2 +
\hbar{\partial_t \theta} + {\hbar^2 \over 2m} 
|{{\boldsymbol\nabla}\theta}|^2)^2 + ... 
\nonumber 
\\
&=& P(\mu) - n \, 
(\hbar {\partial_t \theta} + {\hbar^2 \over 2m} 
|{{\boldsymbol\nabla}\theta}|^2)
+ n {\hbar^2\over 2mc^2} (1+{c^2\over c_s^2}) 
(\partial_t\theta)^2 + ... 
\label{dottica-relaivistic}
\eeqa

\subsection{Still on the penetration depth for the electric field}

On the ground of relativistic invariance, one should expect the penetration lengths for electric and magnetic fields to be the same \cite{sala2023}. However, using the 
nonrelativistic only-phase Popov action 
we have found that the two penetration lengths differ by five orders of magnitude. We now show that by adopting the relativistic only-phase Popov action we still have $\lambda_E\ll \lambda_L$. 

Inserting the electromagnetic potentials of Eqs. (\ref{eanda}) and (\ref{banda}) into  Eq. (\ref{dottica-relaivistic}) but using Eqs. (\ref{schabo1}) and (\ref{schabo2}) we find an extension of Eq. (\ref{gulp1}), namely 
\beq
{\rho} = q \, n - q \bar{n}_{\rm bg} - \epsilon_0 \, {q^2 n \mu_0\over m} 
\left(1+ {c^2\over {  c_s}^2}\right) \, \Phi  
\label{gulp1-relativistic}
\eeq
 It is important to stress that in the relativistic approach of Ref. \cite{sala2023} it was found instead 
 \beq 
\rho = q \, n - q \bar{n}_{\rm bg} - \epsilon_0 \, {q^2 n \mu_0\over m} \, \Phi  \; . 
\eeq
The difference is due to the fact that the relativistic only-phase Popov Lagrangian contains a term  
that is missing in the mean-field treatment of the Klein-Gordon Lagrangian developed in Ref. \cite{sala2023}. 
The crucial point is that the Legendre transformation from the density-phase action to the only-phase action introduces beyond-mean-field contributions. This is indeed the crucial idea, developed in 1972 by Popov \cite{popov-book}, but not yet fully appreciated. In our case, the beyond-mean-field contribution is directly related to the speed of sound $c_s$ of the system. As a consequence of Eq. (\ref{gulp1-relativistic}) the penetration depth of the static electric field reads 
\beq
\lambda_E = \lambda_L {1\over \sqrt{1+{c^2\over c_s^2}}} \simeq \lambda_L {c_s\over c} \;  
\label{lambdaE-relativistic}
\eeq
because $c_s \ll c$ for available superconductors, as discussed in the previous section. 

}

\section{Conclusions} \label{sec:4conclusions}

We have {  corroborated the main idea of Popov, by establishing that} 
the hydrodynamic Lagrangian density of a nonrelativistic 
superfluid, which is 
characterized by the Nambu-Goldstone real scalar field $\theta({\bf r},t)$,  
is obtained with the prescription of Eq. (\ref{prescrivo}) 
into the pressure $P(\mu)$. 
The Lagrangian (\ref{magic0}) is Galilei invariant and it is known as 
the only-phase Popov Lagrangian. Clearly, the real-time action 
functional of Eq. (\ref{final}) is the only-phase Popov 
action we were looking for. 
Our derivation of this nice only-phase action is strictly 
based on the assumption given by Eq. (\ref{assum}), namely that the starting 
density-phase action is exactly Eq. (\ref{initial0}), where 
the quantum-pressure term $\hbar^2(\nabla \sqrt{n})^2/(2m)$ 
has been neglected. In other words, we have demonstrated that ${\tilde S}[n]$ 
of Eqs. (\ref{final}) and (\ref{magic0}) is obtained from $S[n,\theta]$ of 
Eqs. (\ref{initial}) and (\ref{assum}) by performing the 
saddle-point approximation of the functional integration 
with respect to the local density $n$, as shown in Eq. (\ref{mainornot}). 

As an application of the only-phase Popov action, we have 
studied a zero-temperature charged superfluid (superconductor) 
minimally coupled to the electromagnetic field. Notice that our specific application needs the Popov action but not the full theoretical formalism used to get the action. With the help of the only-phase Popov action 
we have obtained quite peculiar dependences of 
charged density and charged current density on the electromagnetic 
scalar and vector potentials. Our findings suggest (see also 
\cite{hirsch2004,hirsch2015,
grigorishin2021}) that, close to zero temperature, there is a strong screening of the electrostatic field 
inside a superconductor with a characteristic length   
much smaller than the London penetration depth. In solids 
$c_s\simeq  10^4$ meters/seconds and consequently we expect $\lambda_E \simeq 10^{-5} \lambda_L$ that is well below the Thomas-Fermi screening length of normal metals. In this paper the model for superconductivity is nonrelativistic, while the fully relativistic model is analyzed in \cite{sala2023}, where the same penetration length is found for both electric and magnetic fields at zero temperature, as expected from the relativistic invariance of the two fields. {\cblue In the last part of this paper we have compared the two theories developing a relativistic only-phase Popov action. This relativistic model contains a beyond-mean-field term, not taken into account in Ref. \cite{sala2023}, that depends on the speed of sound $c_s$ of the system implying $\lambda_E\simeq (c_s/c) \lambda_L$ 
for $c_s/c \ll 1$, with $c$ the speed of light.}
From the experimental point of view, it is not an easy task to measure the penetration depth of the electric field in a superconductor}: in 2016 an attempt on 
a Niobium sample was inconclusive \cite{peronio2016}. 
The main difficulty in the measurement was in combining atomic force microscopy and cryogenic cooling. Unfortunately, the experimental error was so large that no conclusive statement can be made: more accurate experiments at ultra-low temperatures are needed. 

\section*{Acknowledgements}

FL and LS acknowledge a National Grants of the Italian Ministry of 
University and Research: PRIN 2022 project ``Quantum Atomic Mixtures: Droplets, Topological Structures, and Vortices" {  and the project "Eccellenza Dipartimentale - Frontiere Quantistiche"}. LS is partially supported by the BIRD grant ``Ultracold atoms in curved geometries” of the University of Padova, by the “Iniziativa Specifica Quantum” of INFN, by the European Quantum Flagship project PASQuanS 2, and by the European Union-NextGenerationEU within the National Center for HPC, Big Data and Quantum Computing (Project No. CN00000013, CN1 Spoke 10: “Quantum Computing”). 

\section*{Data availability statement}
No new data were created or analysed in this study.

\end{document}